# Vectorial Control of Magnetization by Light

NATSUKI KANDA[1], TAKUYA HIGUCHI[1], HIROKATSU SHIMIZU[1], KUNIAKI KONISHI[1,2], KOSUKE YOSHIOKA[1,2,3] AND MAKOTO KUWATA-GONOKAMI[1,2,3*]

[1]Department of Applied Physics, The University of Tokyo and Core Research for Evolutional Science and Technology (CREST), 113-8656, Tokyo, Japan

[2]Photon Science Center, The University of Tokyo, 113-8656, Tokyo, Japan

[3]Department of Physics, The University of Tokyo, 113-0033, Tokyo, Japan

*e-mail: gonokami@phys.s.u-tokyo.ac.jp

**Coherent light-matter interactions have recently extended their applications to the ultrafast control of magnetization in solids[1-8]. An important but unrealized technique is the manipulation of magnetization vector motion to make it follow an arbitrarily designed multi-dimensional trajectory. Furthermore, for its realization, the phase and amplitude of degenerate modes need to be steered independently. A promising method is to employ Raman-type nonlinear optical processes[2,9-11] induced by femtosecond laser pulses, where magnetic oscillations are induced impulsively with a controlled initial phase and an azimuthal angle that follows well defined selection rules determined by the materials' symmetries[12]. Here, we emphasize the fact that temporal variation of the polarization angle of the laser pulses enables us to distinguish between the two degenerate modes. A full manipulation of two-dimensional magnetic oscillations is demonstrated in antiferromagnetic NiO by employing a pair of polarization-twisted optical pulses. These results have lead to a new concept of vectorial control of magnetization by light.**

Recent advances in ultrashort-pulsed laser technology have enabled ultrafast control of magnetization by light. This has created a new field that is attracting remarkable attention due to both the scientific interest and its potential applications, such as the coherent control of the precession of a single spin[4] or a spin ensemble[2,5-7]. Several experiments revealed a wide variety of physics within this context, for example ultrafast control through a nonlinear optical process[2] or by employing a magnetic component of terahertz (THz) electromagnetic pulses[13,14]. The previous studies mainly focused on the control of the phase and amplitude of a single magnetic oscillation mode. In this case, the three-dimensional trajectory of magnetization motion cannot be altered. As a next step, we propose a new technique to control magnetization in a multidimensional space to fully employ its vectorial properties. This technique can lead to further possibilities such as storing multiple pieces of information in a single storage element and implementation of novel quantum processing using light-spin interactions.

To realise such vectorial control, it is necessary to independently handle phase and amplitude of multiple degenerate modes that constitute complete bases for describing the magnetization vector dynamics, as shown in Fig. 1a. For example, consider a magnetization vector oscillating in a two-dimensional space, whose motion is described by an isotropic harmonic oscillator. The solutions are given by linear combinations of two degenerate orthogonal modes labelled $x$ and $y$ (see Fig. 1b), and the relative phase and amplitude determine its trajectory.

For demonstration of this vectorial control of magnetization, an as-grown single crystal of antiferromagnetic (AFM) NiO (111) which has a micro-multi-domain structure is an excellent candidate. Here, we employed Raman-type nonlinear optical processes for the excitation of magnetic oscillations[2,9-11] and terahertz time-domain spectroscopy (THz-TDS) for detection[15]. This combination of material and techniques has several advantages as described below. First, magnetic susceptibility of this multi-domain sample is given by a scalar quantity because macroscopic averaging of the magnetization over domains is valid, as examined by linear absorption measurements[12]. The susceptibility has a sharp peak in the proximity of the AFM resonance[16], and thus this macroscopic magnetization behaves as an isotropic harmonic oscillator as far as linear optical response is concerned. Second, due to its strong exchange field and high Néel temperature ($T_N$ = 523 K), the resonant frequency $\Omega_{mag}$ is as high as ~1 THz at room temperature. This mode is infrared active[17,18], and this resonant frequency is well within the scope of THz-TDS[12,15,19]. Therefore, we could directly obtain the magnetization trajectory projected on a plane by observing the polarization states of the THz wave radiated from the oscillations through a magnetic dipole radiation process[20]. The electric field $\mathbf{E}(\Omega)$ radiated from the oscillating magnetization $\mathbf{M}(\Omega)$ at a frequency $\Omega$ is described as

$$\mathbf{E}(\Omega) = -(1/\varepsilon\chi_\mu\Omega)\mathbf{k} \times \mathbf{M}(\Omega), \quad (1)$$

where $\varepsilon$, $\chi_\mu$ and $\mathbf{k}$ are the permittiviy, the magnetic linear susceptibility and the wave vector of the radiation, respectively.

Under irradiation by short laser pulses with broadband spectra, magnetization is induced according to the following equation[21]:

$$M_i^{(2)}(\Omega) = \int d\omega \chi_{ijk}^{(2)\text{MEE}}(\Omega;-\omega,\omega+\Omega)E_j^*(\omega)E_k(\omega+\Omega), \quad (2)$$

where $\mathbf{E}(\omega)$ is the Fourier component of the electric field of the excitation optical pulse at frequency $\omega$, $\mathbf{M}^{(2)}(\Omega)$ is the Fourier component of the second-order nonlinear magnetization at a frequency $\Omega$ and $\chi_{ijk}^{(2)\text{MEE}}$ is the third-rank axial time-odd tensor describing the nonlinear susceptibility. In this Raman process, magnetic oscillations are excited impulsively within the pulse duration ~100 fs, which is much shorter than an oscillation cycle. Therefore, by changing the laser pulse timing, the phase of an oscillation mode can be changed, as we demonstrated by coherent control of a magnetic oscillation mode by double-pulse excitation.

Note that this nonlinear optical process does not induce excess heating because light carries away the extra energy which corresponds to the energy mismatch between the magnetic oscillations (~1 meV) and light (~1 eV). They interact with each other coherently, following well-defined polarization selection rules determined by the crystal's symmetry. As stated above, the effective medium approximation is valid for multi-domain NiO. As a result, the system possesses a threefold rotational symmetry around the [111] axis. While this symmetry assures the isotropic linear optical responses, the nonlinear optical response is not isotropic[22], and the direction of the excited magnetic oscillations is determined by the polarization angle of the excitation laser pulse[12]. Thus, the two degenerate modes can be controlled independently as we will discuss below in detail.

Figure 1 shows a schematic depiction of the experiment. Ti:Sapphire-based femtosecond laser pulses were employed as the pumping beam, which propagated along the [111] axis of an as-grown bulk single crystal of NiO. The pump pulses were separated into a pair of pulses in which the pulse interval and polarization angles were under control. The polarization properties of the THz electromagnetic waves radiated from the induced dynamic magnetization were studied with the electro-optic sampling method[15] and the THz polarimetry technique using a pair of wire-grid polarizers[23]. (See Methods Summary for details of the experimental setup.) All measurements were performed at room temperature.

Employing double-pulse excitation, we perform coherent control[24,25] of a magnetic oscillation mode as shown in Fig. 1a in which the polarization angles of the two pulses are the same (see Fig. 2a). The decay time of this magnetic resonance was on the order of tens of picoseconds[12,19], which was much longer than a cycle of the magnetic oscillations. Within this coherence time, we could superpose the magnetic oscillations induced by the first and second pulses. The amplitude of the THz signal after the excitation by the second pulse changed periodically as a function of the time interval $\tau$ between the pulses (see Fig. 2b). Because of the monochromatic nature of the AFM resonance, $\Omega_{mag}\tau$ determined the phase difference of the magnetic oscillations induced by the two optical pulses. When $\Omega_{mag}\tau/\pi$ was an even integer, they constructively interfered, resulting in an enhancement of its amplitude. In contrast, when $\Omega_{\mathrm{mag}}\tau/\pi$ was an odd integer, they destructively interfered, and the magnetization was cancelled.

We extended this technique to arbitrary control of the trajectory of the magnetic oscillations, namely independent control of the two degenerate modes. Before discussing this experiment, we should note an important point that such arbitrary control could not be realized by a single excitation pulse with a fixed polarization, although a plane wave of light has the same number of parameters as that of the two-dimensional motion of magnetization. This is because the information of the carrier phase of the light was lost in the Raman process, decreasing the number of control parameters.

In the case of multi-domain NiO (111), the non-vanishing transverse components of the nonlinear susceptibility tensor $\chi_{ijk}^{(2)\mathrm{MEE}}(\Omega; -\omega, \omega+\Omega)$ are $\chi_{yxx}^{(2)\mathrm{MEE}} = -\chi_{yyy}^{(2)\mathrm{MEE}} = \chi_{xxy}^{(2)\mathrm{MEE}} = \chi_{xyx}^{(2)\mathrm{MEE}} \equiv \alpha(\Omega, \omega)$. Equation (2) is reduced to the following:

$$\begin{bmatrix} M_x^{(2)}(\Omega) \\ M_y^{(2)}(\Omega) \end{bmatrix} = \int d\omega\, \alpha(\Omega,\omega) \begin{bmatrix} E_x^*(\omega)E_y(\omega+\Omega) + E_y^*(\omega)E_x(\omega+\Omega) \\ E_x^*(\omega)E_x(\omega+\Omega) - E_y^*(\omega)E_y(\omega+\Omega) \end{bmatrix}. \quad (3)$$

When the incident laser has a fixed polarization, i.e. the ratio between the complex amplitudes of the electric fields $E_x(\omega)$: $E_y(\omega)$ does not depend on $\omega$, $M_x$ and $M_y$ are always real-valued quantities, according to this equation. In other words, one cannot tune the relative phase of the two orthogonal oscillations. The result is that the THz wave radiated from $\mathbf{M}^{(2)}$ is always linearly polarized. On the other hand, the direction of the oscillation can be tuned by changing the polarization azimuthal angle $\phi$ of the incident laser with respect to the $x$-axis ([−110] axis of the crystal), where $E_x(\omega) = E(\omega)\cos\phi$, $E_y(\omega) = E(\omega)\sin\phi$, and Eq. (3) is reduced to

$$\begin{bmatrix} M_x^{(2)}(\Omega) \\ M_y^{(2)}(\Omega) \end{bmatrix} = \int d\omega\, \alpha(\Omega,\omega) E^*(\omega) E(\omega+\Omega) \begin{bmatrix} \sin 2\phi \\ \cos 2\phi \end{bmatrix}. \quad (4)$$

To handle the relative phase of the two basis modes, we employed a pair of linearly polarized laser pulses with different polarization angles. As determined from Eq. (4), the azimuthal angle of the induced magnetic oscillation changed by $-2\Delta\phi$ when $\phi$ changed by $\Delta\phi$. This polarization angle dependence was generally valid when the system had a threefold rotational symmetry. Therefore, two orthogonal oscillation modes can be accessed independently if the angle between the azimuths of the two pulses is 45° (See Fig. 3a). By changing the interval between these two pulses, the relative phase of the oscillation modes is controlled because the oscillation is kicked instantaneously within the ultrashort duration of a laser pulse. Figure 3b shows the measured ellipticity $(|M_r| - |M_l|)/(|M_r| + |M_l|)$ as a function of the delay time $\tau$, where $M_r$ and $M_l$ are the right and left circular components of **M**, respectively. The polarization of the THz wave radiated from **M** was linear when $\Omega_{mag}\tau/\pi$ was an integer and was circular when it was a half-integer. We continuously changed the ellipticity from $-1$ to $+1$ by sweeping $\tau$. Note that selective excitation of circularly polarized magnetization is possible independent of the crystal azimuthal orientation.

We demonstrate a method for all-optical manipulation of the dynamics of a magnetization vector in NiO. By designing timing- and polarization-tuned double-pulse excitation, selective manipulation of two degenerate oscillation basis modes was achieved based on well-defined polarization selection rules. In particular, simply changing the interval between two laser pulses selectively induced clockwise or anticlockwise rotational motion of the magnetization vector. The scope of our technique is not limited to this specific example since selective control of rotation means that angular momentum transferred from the light to the material is tuned. For example, initialization of a single spin and selective manipulation of degenerate elementary excitations are promising possible applications. Here, temporal variation of the polarization angle of the incident light plays a key role in this technique. Thus, an approach employing polarization-shaped laser pulses[26] should be further examined. In addition to the applications in spintronics, this technique has potential in THz technology in which controlling elementary excitations that fall in the THz spectral range is of particular interest.

**Acknowledgments**

We are grateful to Yuki Shiomi for sample evaluations by X-ray diffraction and Yuri P. Svirko, Hiroharu Tamaru, Jean Benoit Héroux and Toshihiko Shimasaki for their fruitful discussions. This research was supported by the Photon Frontier Network Program, Global COE Program 'the Physical Sciences Frontier', Special Coordination Funds for Promoting Science and Technology of the Ministry of Education, Culture, Sports, Science and Technology, Japan, KAKENHI (20104002) and Research Fellowships for Young Scientists (N. K. and T. H.) from the Japan Society for the Promotion of Science.

**Methods Summary**

**Materials**

An as-grown single crystal of NiO (111) with a thickness of 100 μm was used for this study. The direction of spin ordering and associating lattice distortions resulted in 12 kinds of domains in this crystal, which were randomly distributed in the as-grown sample with the same population, as described below. NiO crystallizes in a rocksalt structure, and below its Néel temperature $T_N$ = 523 K, the spins are ordered ferromagnetically in the {111} plane, and alternate stacking of these layers forms an antiferromagnetic (AFM) structure[27]. Due to this magnetic ordering, slight rhombohedral distortions were induced along the diagonals, [111] and three other equivalents, which were accommodated by twin structures, forming four kinds of T-domains. Experimentally, the T-domains were distinguished by the birefringence due to their lattice distortions. The sizes of the T-domains in our sample were smaller than 10 μm, according to polarization microscopy. In each T-domain, there were three equivalent spin orientations. They formed three kinds of S-domains, and thus there were 12 kinds of domains in total. For example, in the $T_1$-domain, in which spins were ordered in the (111) plane, the spins aligned along [11-2], [1-21] or [-211]. These S-domains were also randomly distributed, and they were typically smaller than 1 μm[28].

In this micro-multi-domain sample, a sharp AFM resonance in the vicinity of ~1 THz was observed by absorption measurements employing THz-TDS. No other resonance was observed in the spectral range of 0.5–2 THz. The linear absorption showed no dependence on the polarization of the incident THz wave—i.e. there was neither birefringence nor chirality in the THz range.

In our experimental conditions, the effects of the phase mismatch were negligible because the coherent length of the nonlinear optical interaction was 130 μm, which exceeded the sample thickness (100 μm), since the refractive indices of NiO for the THz wave and fundamental excitation light were 3.5 and 2.3, respectively[29]. Note that the absorption coefficient of NiO for a wavelength of 800 nm was small[30], and thus we could eliminate indirect magnon generation processes through one-photon absorption followed by thermal relaxation processes.

**Experimental setup**

A Ti:Sapphire-based regenerative amplifier was used as a light source. The central wavelength was 800 nm, the pulse duration was 140 fs and the repetition rate was 1 kHz. The beam was divided into two beams and used as a pump beam for sample excitation and a probe beam for electro-optic (EO) sampling. The spot size of the excitation on the sample was approximately 1 mm in diameter, which is much larger than the single domain size (~10 μm), and thus coherent superposition of the signals generated in each of the domains was valid. The pulse energy was less than ~300 μJ. The THz wave radiation was collected with parabolic mirrors, and the electric field was detected via the EO sampling method using a ZnTe (110) crystal with a thickness of 1 mm. The electric field vector of the THz radiation was obtained via THz polarimetry using two wire-grid polarizers[23].

**References**

1. Kirilyuk, A., Kimel, A. V. & Rasing, T. Ultrafast optical manipulation of magnetic order. *Rev. Mod. Phys.* **82**, 2731-2784 (2010).

2. Kimel, A. V., Kirilyuk, A., Usachev, P. A., Pisarev, R. V., Balbashov, A. M. & Rasing, T. Ultrafast non-thermal control of magnetization by instantaneous photomagnetic pulses. *Nature* **435**, 655-657 (2005).

3. Duong, N. P., Satoh, T. & Fiebig, M. Ultrafast Manipulation of Antiferromagnetism of NiO. *Phys. Rev. Lett.* **93**, 117402 (2004).

4. Kikkawa, J. M. & Awschalom D. D. Resonant Spin Amplification in n-Type GaAs. *Phys. Rev. Lett.* **80**, 4313–4316 (1998).

5. Hansteen, F., Kimel, A., Kirilyuk, A. & Rasing, T. Nonthermal ultrafast optical control of the magnetization in garnet films. *Phys. Rev. B* **73**, 014421 (2006).

6. Bigot, J.-Y., Vomir, M. & Beaurepaire E. Coherent ultrafast magnetism induced by femtosecond laser pulses. *Nature Physics* **5**, 515-520 (2009).

7. Hanson, R. & Awschalom, D. Coherent manipulation of single spins in semiconductors. *Nature* **453**, 1043-1049 (2008).

8. Satoh, T., Cho, S.-J., Iida, R., Shimura, T., Kuroda, K., Ueda, H., Ueda, Y., Ivanov, B. A., Nori, F. & Fiebig M. Spin Oscillations in Antiferromagnetic NiO Triggered by Circularly Polarized Light. *Phys. Rev. Lett.* **105**, 077402 (2010).

9. Lockwood, D. J., Cottam, M. G. & Baskey, J. H. One- and two-magnon excitations in NiO. *J. Magn. Mag. Mater.* **104**, 1053-1054 (1992).

10. Fleury, P. A. & Loudon, R. Scattering of Light by One- and Two-Magnon Excitation. *Phys. Rev.* **166**, 514-530 (1968).

11. Shen, Y. R. & Bloembergen, N. Interaction between Light Waves and Spin Waves. *Phys. Rev.* **143**, 372-384 (1966).

12. Higuchi, T., Kanda, N., Tamaru, H. & Kuwata-Gonokami, M. Selection Rules for Light-Induced Magnetization of a Crystal with Threefold Symmetry: The Case of Antiferromagnetic NiO. *Phys. Rev. Lett.* **106**, 047401 (2011).

13. Kampfrath, T., Sell, A., Eilers, G., Wolf, M., Fiebig, M., Leitenstorfer, A., Münzenberg, M. & Huber, R. Coherent terahertz control of antiferromagnetic spin waves. *Nature Photon.* **5**, 31-34 (2011).

14. Nakajima, M., Namai, A., Ohkoshi, S. & Suemoto, T. Ultrafast time domain demonstration of bulk magnetization precession at zero magnetic field ferromagnetic resonance induced by terahertz magnetic field. *Opt. Express* **18**, 18260-18268 (2010).

15. Wu, Q. & Zhang, X. C. Ultrafast electro-optic field sensors. *Appl. Phys. Lett.* **68**, 1604-1606 (1996).

16. Kittel, C. Theory of Antiferromagnetic Resonance. *Phys. Rev.* **82**, 565 (1951).

17. Kondoh, H. Antiferromagnetic Resonance in NiO in Far-infrared Region. *J. Phys. Soc. Jpn.* **15**, 1970-1975 (1960).

18. Sievers, A. J. & Tinkham, M. Far Infrared Antiferromagnetic Resonance in MnO and NiO. *Phys. Rev.* **129**, 1566-1571 (1963).

19. Nishitani, J., Kozuki, K., Nagashima, T. & Hangyo M. Terahertz radiation from coherent antiferromagnetic magnons excited by femtosecond laser pulses. *Appl. Phys. Lett.* **96**, 221906 (2010).

20. Jackson, J. D. *Classical Electrodynamics*, 2nd Edition, edited John Wiley & Sons, (New York, 1975).

21. Pershan, P. S. Nonlinear Optical Properties of Solids: Energy Considerations. *Phys. Rev.* **130**, 919–929 (1963).

22. Bloembergen, N. Conservation laws in nonlinear optics. *J. Opt. Soc. Am.* **70**, 1429-1436 (1980).

23. Kanda, N., Konishi, K. & Kuwata-Gonokami, M. Terahertz wave polarization rotation with double layered metal grating of complimentary chiral patterns. *Opt. Express* **15**, 11117-11125 (2007).

24. Dekorsy, T., Kutt, W., Pfeifer, T. & Kurz, H. Coherent Control of LO-Phonon Dynamics in Opaque Semiconductors by Femtosecond Laser Pulses. *Europhys. Lett.* **23**, 223-228 (1993).

25. Planken, P. C. M., Brener, I., Nuss, M. C., Luo, M. S. C. & Chuang, S. L. Coherent control of terahertz charge oscillations in a coupled quantum well using phase-locked optical pulses. *Phys. Rev. B* **48**, 4903-4906 (1993).

26. Brixner, T. & Gerber, G. Femtosecond polarization pulse shaping. *Opt. Lett*. **26**, 557-559 (2001).

27. Roth, W. L. Magnetic Structures of MnO, FeO, CoO, and NiO. *Phys. Rev.* **110**, 1333-1341 (1958).

28. Sänger, I., Pavlov, V. V., Bayer, M. & Fiebig, M. Distribution of antiferromagnetic spin and twin domains in NiO. *Phys. Rev. B* **74**, 144401/1-9 (2006).

29. Powell, P. J. & Spicer, W. E. Optical Properties of NiO and CoO. *Phys. Rev. B* **2**, 2182-2193 (1970).

30. Newman, R. & Chrenko, R. M. Optical Properties of Nickel Oxide. *Phys. Rev.* **114**, 1507-1513 (1959).

**Figure legends**

Figure 1. **Schematic illustrations of the experiment. a.** Vectorial control of the magnetic oscillations and the resultant THz radiation by double-pulse excitation. Precession motions of the spins are induced in a multi-domain single crystal of NiO by the stimulated Raman-type nonlinear optical process, as shown in the inset. Spin motions result in a macroscopic magnetization vector ($\mathbf{M}^{(2)}$) which behaves as an isotropic harmonic oscillator. By changing the interval and polarization of the incident laser pulses, an arbitrary trajectory of the magnetic oscillations is obtained. **b.** Two orthogonal magnetic oscillation modes labelled by $x$ and $y$ in this NiO crystal. They are selectively kicked by tuning the polarization azimuth of the excitation light pulse.

Figure 2. **Coherent control of the oscillations of magnetization in NiO using linearly polarized double-pulse excitation with the same polarization angle. a.** Schematic illustrations of the experiment of coherent control with double-pulse excitation in which both pulses are linearly polarized along the $x$-axis. The interval $\tau$ between the two pulses can be tuned. **b**. Two-dimensional (2D) plot of the time-domain THz signal. The periodically aligned horizontal white lines correspond to the suppression of the spin motion by the destructive interference between two waves induced by the two excitation pulses. **c**. Cross sections of the 2D plot with fixed $\tau$ in which the two waves interfere with each other (i) constructively or (ii) destructively.

Figure 3. **Vectorial control of the magnetization with polarization-twisted double-pulse excitation. a.** Schematic illustrations of vectorial control of the magnetization vector. Both pulses are linearly polarized. The first one is polarized along the $x$-axis and the second one has an azimuthal angle of $\psi$. By properly tuning $\tau$ and $\psi$, we can manipulate the motion of the magnetization vector to follow an arbitrarily designed direction and amplitude of polarization.

**b.** Ellipticity of the measured THz radiation at $\Omega_{mag}$ as a function of the interval $\tau$ between the two linearly polarized excitation pulses. The first pulse is *x*-polarized and the polarization azimuth of the second is 45° with respect to the *x*-axis. Any ellipticity between the purely right and left circular polarizations is obtained by tuning $\tau$. The solid curve is a fit of the data by a sine curve. The lower panels show the three-dimensional trajectories of the electric field vectors with fixed $\tau$ in which the radiation is (i) linearly polarized or (ii) purely circularly polarized. The dots are the projections of the experimental data onto the *x*- and *y*-axes. The solid curves are the fit of exponentially decaying sine functions; the phase differences between the *x*- and *y*-components are determined by $\Omega_{mag}\tau/\pi$.

**Fig. 1**

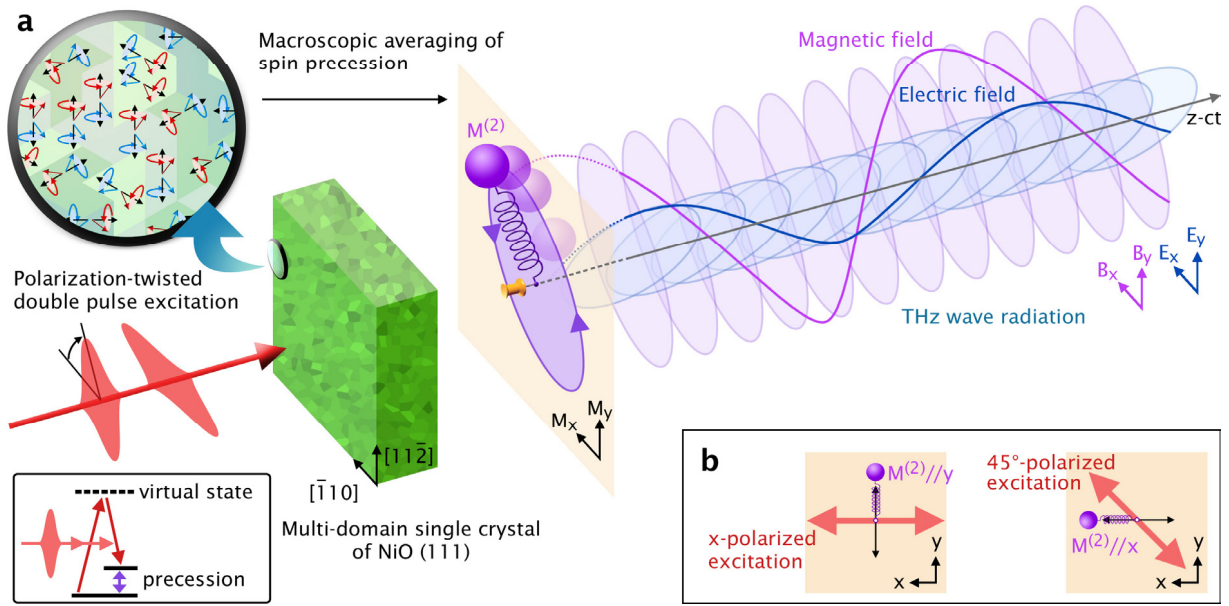

**N. Kanda**

**Fig. 2**

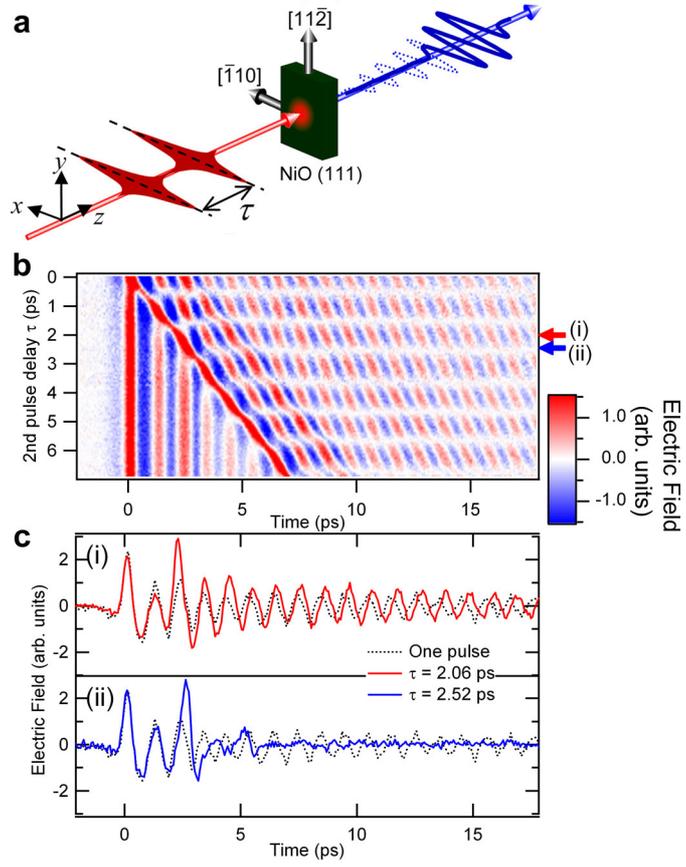

N. Kanda

**Fig. 3**

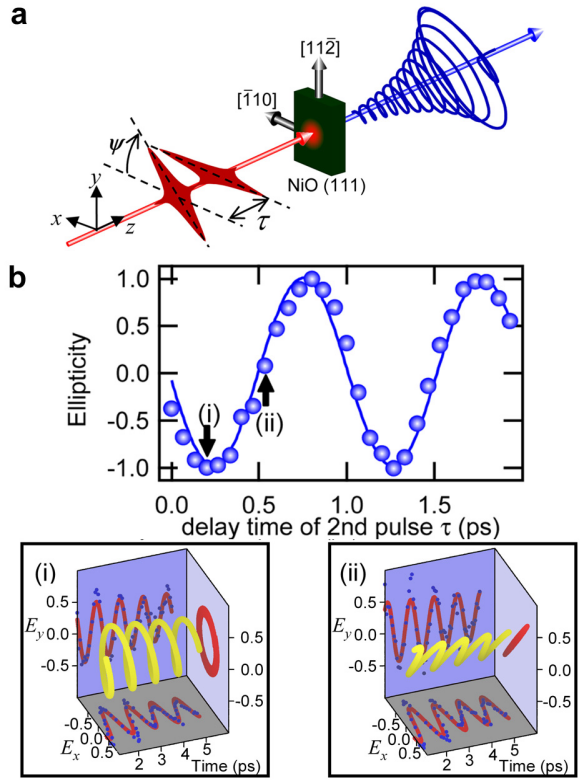

N. Kanda

17